\documentclass[12pt,a4paper,reqno]{amsart}
\usepackage[latin1]{inputenc}  
\usepackage[T1]{fontenc}       
\usepackage{amsfonts}          
\usepackage{amsmath}
\usepackage{amssymb}
\usepackage{graphicx}
\usepackage{url}
\newcommand{\R}{\mathbb{R}}

\newcommand{\Z}{\mathbb{Z}}

\def\viiva #1 #2 {\mathop{\Big/}\limits_{\!\!\!{#1}}^{\>\,{#2}}}

\numberwithin{equation}{section}

\begin{document}

\title{Exact asymptotics of the freezing transition of a logarithmically correlated random energy model}
\author{Christian Webb\\ University of Helsinki, Department of Mathematics\\
christian.webb@helsinki.fi}
\thanks{This research was funded by the Academy of Finland}

\maketitle

\subsection*{Abstract}

\vspace{0.3cm}

We consider a logarithmically correlated random energy model, namely a model for directed polymers on a Cayley tree, which was introduced by Derrida and Spohn. We prove asymptotic properties of a generating function of the partition function of the model by studying a discrete time analogy of the KPP-equation - thus translating Bramson's work on the KPP-equation into a discrete time case. We also discuss connections to extreme value statistics of a branching random walk and a rescaled multiplicative cascade measure beyond the critical point.

\section{Introduction}

In \cite{DS} Derrida and Spohn introduced a model for a directed polymer on a disordered Cayley tree. They put i.i.d. random potentials on each bond of the tree, considered self avoiding walks $\omega$ on the tree and the partition function $\mathcal{Z}(t)=\sum_{\omega}e^{-\beta E(\omega)}$, where the sum is over the paths $\omega$ which are self avoiding and of length $t$, $\beta$ is the inverse temperature and $E(\omega)$ is the sum of the potentials at each bond the walk crosses. They then argued that a suitable generating function for $\mathcal{Z}$, ($G_t(x)=\mathbb{E}(\exp(-e^{-\beta x}\mathcal{Z}(t)))$) satisfies the following non-linear integro-difference equation 

\begin{equation}\label{eq:recursion}
G_{t+1}(x)=\int \rho(y) G_t(x+y)^Kdy,
\end{equation}

\vspace{0.3cm}

\noindent with initial data $G_0(x)=\exp(-e^{-\beta x})$. Here $\rho$ is the density of the distribution of the random potentials and $K$ is the order of the tree: each site apart from the root (which has a single neighbor) has $K+1$ neighbors. 

\vspace{0.3cm}

The main physical interest in this model is that it is simple enough that one can analyze it in a fair amount of detail, but it is also rich enough to contain non-trivial logarithmic correlations between the energies of the paths and hopefully some universal properties of more complicated disordered systems with such correlations. Another way to describe this model is that it is a random energy model with logarithmic correlations. As anonther reference to the model, we direct the reader to \cite{Fyodorov}, where the relationship between random multifractal measures and logarithmically correlated random energy models are discussed. In particular, it is explained here why the model is logarithmically correlated.

\vspace{0.3cm}

Instead of analyzing this model in great depth, Derrida and Spohn conjectured that the system behaves similarly to a continuum one, where the walks are replaced by Brownian motion and the splitting in the tree happens at a random exponentially distributed time. They then argued that for $K=2$, the corresponding quantity $G_t(x)$ satisfies the KPP-equation

\begin{equation}\label{eq:kpp}
\partial_t G_t(x)=\frac{1}{2}\partial_x^2 G_t(x)+G_t(x)^2-G_t(x).
\end{equation}

\vspace{0.3cm}

This equation has been studied extensively by Bramson \cite{Bramson}. His results imply that there is a function $m^\beta(t)$ so that $G_t(x+m^\beta(t))$ converges to a traveling wave solution $g_\beta$ (i.e., for a certain $c(\beta)$,  $g_\beta(x-c(\beta)t)$ is a solution to \eqref{eq:kpp}). One particular phenomenon emerging from Bramson's analysis is that there is a phase transition in the system. At a certain critical temperature the system freezes. This can be seen for example from the form of  $c(\beta)$ and $g_\beta(x)$ as well as the asymptotics of $m^\beta(t)$: these all become independent of $\beta$ for large enough $\beta$.

\vspace{0.3cm}

This freezing seems to be a phenomenon occurring in a wide range of disordered systems (see \cite{FB} for a further discussion and references). Indeed some non-rigorous work by Carpentier and Le Doussal \cite{CLD} suggests that this freezing is something occurring quite generally in systems with logarithmically correlated disorder.

\vspace{0.3cm}

The discrete model is also related to many purely mathematical questions. For example, at zero temperature, only the lowest energy configuration is relevant. So the $\beta=\infty$ case is closely related to the question of extreme value statistics, i.e., finding the distribution of $\min\lbrace V_1,...,V_N\rbrace$ in the $N\rightarrow \infty$ limit, where the $V_i$ have logarithmic correlations described by the model. In the case where there are no correlations, this distribution is well known and it is known to extend to a large class of correlated random variables. Our analysis will imply that the correlations in our model are beyond this universality class. In \cite{FB}, there is an interesting conjecture about the exact form of the extreme value statistics of a certain logarithmically correlated system. 

\vspace{0.3cm}

Instead of considering each self avoiding walk on the tree separately, one can consider them to be a single branching random walk. This allows one to define some useful martingales. Branching random walks have been considered more generally and in great detail by Biggins and others (see e.g. \cite{Biggins}). The $\beta=\infty$ case and the problem of extreme value statistics can be interpreted as the problem of finding the distribution of the minimum of a branching random walk (or in our case equivalently the maximum since we shall be dealing with symmetric random variables). This is a problem that has been studied recently for quite general branching random walks (see e.g. \cite{BZ} and \cite{ABR}). Moreover, this interpretation of the maximum of a branching random walk can be used to study the maximum of the discrete two dimensional Gaussian Free Field (see \cite{BDZ} and \cite{BZ2}).

\vspace{0.3cm}

Another interesting problem related to the model is that one can use the energies of the paths to construct random measures on hypercubes - so called multiplicative cascade measures (see \cite{KP} and \cite{BS} for information about multiplicative cascades). For $K=2^n$, one splits the $n$-dimensional unit hypercube into $K^s$ equal sized hypercubes at stage $s$. At stage $t$ one gives a hypercube the weight $e^{-\beta E_i-\frac{\beta^2}{2}t}$ where $E_i$ is the energy of a path one identifies with a sequence of nested hypercubes and the term $\frac{\beta^2}{2}t$ is there so that the expected volume of the unit hypercube is $1$. Several things are known about such measures (\cite{KP,BS}). First of all, a weak limit (as $t\rightarrow \infty$) exists almost surely. The limit measure has positive total mass if and only if $\beta<\sqrt{2\log K}$ and it has no atoms almost surely. An interesting question is if we can modify the measure in some simple way so that a limit would exist also in the $\beta\geq \sqrt{2\log K}$ case. Moreover, if the limit exists, does it have atoms? These questions are closely related to the notion of multifractality discussed in \cite{Fyodorov}. As \cite{BZ2} suggests a relationship with branching random walks and the discrete two dimensional Gaussian Free Field, one might suspect that these multiplicative cascade measures (and their modified versions) are related to the measures of quantum gravity considered in \cite{SD}. 

\vspace{0.3cm}

Our primary goal will be to show that in the case that the potentials are standard Gaussians, the discrete case indeed behaves as expected, i.e., there is a function $m^\beta(t)$ so that $G_t(x+m^\beta(t))$ converges uniformly to a function which is a solution to a stationary version of the recursion relation \eqref{eq:recursion}. As in the continuum case, the shift needed to obtain a non-trivial limit is given by $m^\beta(t)=G_t^{-1}(\frac{1}{2})$. The stationary version of the recursion relation will turn out to be

\begin{equation}\label{eq:stationary}
w(x)=\int_{-\infty}^\infty \frac{1}{\sqrt{2\pi}}e^{-\frac{1}{2}y^2}w(x+y+c(\beta))^K dy,
\end{equation}

\vspace{0.3cm}

\noindent where

\vspace{0.3cm}

\begin{equation}\label{eq:speed}
c(\beta)=\left\lbrace \begin{array}{l}
\frac{\beta}{2}+\frac{\log K}{\beta},\quad \ \ \ \mathrm{for} \ \beta<\sqrt{2\log K}  \\ 
\sqrt{2\log K}, \quad\quad \mathrm{for} \ \beta\geq\sqrt{2\log K}
\end{array} \right. 
\end{equation}

\vspace{0.3cm}

\noindent We shall also call $w$ a traveling wave, $c(\beta)$ its speed and the equation it satisfies the stationary equation. 

\vspace{0.3cm}

As noted in \cite{FB}, from the point of view of studying universality classes of extreme value statistics, the asymptotic behavior of $w$ is important. In our case,

\begin{equation}\label{eq:hightemptail}
\lim_{x\rightarrow\infty}e^{\beta x} (1-w(x))=C
\end{equation}

\vspace{0.3cm}

\noindent for some $C>0$ and for $\beta\geq \sqrt{2\log K}$, 

\begin{equation}\label{eq:lowtemptail}
\lim_{x\rightarrow\infty}\frac{1}{x}e^{\sqrt{2\log K} x} (1-w(x))=C'
\end{equation}

\vspace{0.3cm}

\noindent for some $C'>0$. Moreover, we shall show that under certain restrictions of the initial data, the solution to the stationary equation is unique up to translations.

\vspace{0.3cm}

We note that the stationary equation enjoys a certain high-temperature self-duality. In the high-temperature regime, the equation is of the form

\begin{equation*}
w(x)=\int_{-\infty}^\infty \frac{1}{\sqrt{2\pi}}e^{-\frac{1}{2}y^2}w\left(x+y+\frac{\beta}{2}+\frac{\log K}{\beta}\right)^K dy.
\end{equation*}

\vspace{0.3cm}

\noindent This equation is clearly invariant under the mapping $\beta\to\frac{\beta_c^2}{\beta}=\frac{2\log K}{\beta}$. While there certainly is no duality between the physics of the high- and low-temperature regimes, this curious formal duality is suspected to be related to physical properties of the model. Indeed, in \cite{FDR} it was noticed that this type of high-temperature self-duality occurs in some more complicated logarithmically correlated random energy models (where verifying may not be quite a s simple) and it was conjectured that this duality property is intimately related to the freezing occurring. Some further support for this conjecture was found in \cite{FDR2}.

\vspace{0.3cm}

From the point of view of the convergence of the random measure mentioned above, one of the first questions to ask would be, does $A_t \mathcal{Z}(t)$ converge, where $A_t$ is some deterministic normalization. Since $G_t(x+m^\beta(t))=\mathbb{E}(\exp(e^{-\beta(x+m^\beta(t))}\mathcal{Z}(t)))$ converges, the only possible normalization would be asymptotically a multiple of $e^{-\beta m^\beta(t)}$. For this, the precise form of $m^\beta(t)$ may be important. This could also be relevant for studying the 2-dimensional discrete Gaussian Free Field since as seen in \cite{BZ2}, $m^\infty(t)$ is related to the expectation of the maximum of the two dimensional discrete Gaussian Free Field. Our result for the asymptotic form of $m^\beta(t)$ is

\begin{equation}\label{eq:mb}
m^\beta(t)=\left\lbrace \begin{array}{l}
c(\beta)t+\mathcal{O}(1),\qquad\qquad\qquad\qquad\ \ \ \ \mathrm{for} \ \beta<\sqrt{2\log K} \\ 
\sqrt{2\log K}t-\frac{1}{2\sqrt{2\log K}}\log t+\mathcal{O}(1)\quad \mathrm{for}\ \beta=\sqrt{2\log K}\\ 
\sqrt{2\log K}t-\frac{3}{2\sqrt{2\log K}}\log t+\mathcal{O}(1)\quad \mathrm{for}\ \beta>\sqrt{2\log K}
\end{array} \right. .
\end{equation}

\vspace{0.3cm}

While it is intuitively rather clear that the discrete system should behave as the continuum one and in the $\beta=\infty$ case this has been showed even for more general non-linearities \cite{BZ,ABR}, a written argument in the generality we are considering seems to be missing. Reading through Bramson's work on the continuum case, one notices that many of the arguments he uses work in the discrete case with minor modifications. Indeed for the part on $m^\beta(t)$, we shall not go over all of the technical details that would be formally identical to those found in \cite{Bramson}, but we shall reproduce the main argument in the discrete time language and provide the additional results that might not be immediately obvious. Moreover, we shall prove the convergence of $G_t(x+m^\beta(t))$ independently of knowing the precise asymptotic behavior of $m^\beta(t)$. This along with the rather small class of initial data we are interested allows one to cut a few corners in following Bramson's reasoning. 

\vspace{0.3cm}

Our proof for the convergence of $G_t(x+m^\beta(t))$ will be rather different from Bramson's for $\sqrt{2\log K}\leq\beta<\infty$ and we shall make use of arguments used in various areas concerning similar problems in discrete and continuous time. In fact, a secondary goal of this note is to collect different kinds of arguments and references to various areas which seem to have been independently working on similar problems with different kinds of approaches. The point of this being that to study more difficult problems such as the existence of the limit measure discussed above, the distribution of the partition function of the two dimensional Gaussian Free Field or problems related to quantum gravity, a wider range of tools could be useful.

\vspace{0.3cm}

When proving convergence, one of our main tools will be a generalized maximum principle type result which is a discrete version of one used by Bramson. Some of our arguments will follow Bramson's approach and some follow \cite{Lui}, where a similar recursion relation is studied, but the density of the random variables has compact support. Another important tool we shall need is a family of martingales related to the branching random walk (see e.g. \cite{Biggins} for a more general discussion of such martingales).  Their properties in the framework of branching diffusions have been studied in \cite{Neveu}, which was applied to the KPP-equation in \cite{Harris}. For the asymptotic behavior of the solution of the stationary equation, we shall rely on work by Durrett and Liggett \cite{DL}. The study of $m^\beta(t)$ follows the work of Bramson closely and our main tool will be a discrete time Feynman-Kac formula and the analysis of a discrete time Brownian bridge.

\vspace{1cm}

\section{Tools for demonstrating convergence: a branching random walk and a maximum principle.}

In this section, we shall go over some basic results related to a branching random walk and a generalized maximum principle for a certain class of integral operators. A lot of the results related to the branching random walk have been found for more general branching random walks by Biggins and others (see e.g. \cite{Biggins}). As already mentioned, our discussion about the branching random walk will rely on work in \cite{Neveu} and \cite{Harris}.

\vspace{0.3cm}

The branching random walk we are interested in is defined in the following manner. We start with a particle located at some position $x$. This particle takes a random step to $x+V$, where $V$ is normally distributed with zero mean and unit variance. After this step, the particle splits into $K$ new particles ($K$ is fixed) all of which are located at $x+V$. After the splitting, one unit of time has elapsed. Each of these particles then behaves as the initial one and independent of the others. So at time $t$ (an integer), we have $K^t$ particles. They are grouped into $K^{t-1}$ clusters of $K$ particles. Let us write $X_k(t)$, $k=1,...,K^t$ for the locations of the $K^t$ particles at time $t$. The indexing is so that $X_1(t),...,X_K(t)$ are in the same cluster, $X_{K+1}(t),...,X_{2K}(t)$ are in the same cluster and so on.

\vspace{0.3cm}

The fundamental objects we shall use are the random variables $Z_\beta(t)=\sum_{k=1}^{K^t} e^{-\beta(X_k(t)+c(\beta)t)}$. The self-similar structure of the branching random walk gives a useful decomposition: $Z_\beta(t+s)=\sum_{k=1}^{K^t}e^{-\beta(X_k(t)+c(\beta)t)}Z_\beta^{(k)}(s,t)$. Here for each $t$, $Z_\beta^{(k)}(s,t)$ are independent copies of $Z_\beta(s)$ and the corresponding branching random walks start from the origin (implying $Z_\beta^{(k)}(0,t)=1$). They are also independent of the process up to time $t$. Let us write $\lbrace\mathcal{F}_t\rbrace_t$ for the filtration of the branching random walk.

\vspace{0.3cm}

The following results concerning the branching random walk are either direct calculations or their proofs are simple modifications of those found in \cite{Neveu} and \cite{Champ}.

\vspace{0.3cm}

\bf Lemma 2.1. \rm For $\beta\leq \sqrt{2\log K}$, $Z_\beta$, $\partial_\beta Z_\beta$ and $\partial_\beta^2 Z_\beta$ are martingales with respect to the filtration of the branching random walk.

\vspace{0.3cm}

\bf Lemma 2.2. \rm For $\beta<\sqrt{2\log K}$, $Z_\beta$ is uniformly integrable.

\vspace{0.3cm}

\bf Lemma 2.3. \rm Let $L(t)=\min_k X_k(t)$. Then $\frac{L(t)}{t}\rightarrow -\sqrt{2\log K}$ almost surely as $t\rightarrow \infty$. 

\vspace{0.3cm}

\bf Lemma 2.4. \rm $\lim_{t\rightarrow \infty} (L(t)+\sqrt{2\log K}t)=\infty$ almost surely. 

\vspace{0.3cm}

Let us write $\rho(y)=\frac{1}{\sqrt{2\pi}}e^{-\frac{1}{2}y^2}$. To prove convergence in the $\beta\geq \sqrt{2\log K}$ case, we shall need the following generalized maximum principle. 

\vspace{0.3cm}

\bf Lemma 2.5. \rm Let $G_t^{1}$ and $G_t^{2}$ be given by the recursion relation \eqref{eq:recursion} and have initial data $G_0^{1}$ and $G_0^{2}$ (measurable and between zero and one). Let us assume that the initial data has the following property: there is a point $x_0\in \R$ so that $G_0^2(x)>G_0^1(x)$ for $x> x_0$ and $G_0^2(x)<G_0^1(x)$ for $x<x_0$. Then for all $t\geq 1$, there is a point $x_t\in[-\infty,\infty]$ so that $G_t^2(x)>G_t^1(x)$ for $x> x_t$ and $G_t^2(x)<G_t^1(x)$ for $x<x_t$. Moreover, if $x_t\in\lbrace -\infty,\infty\rbrace$ for some $t$, then $x_s=x_t$ for all $s\geq t$. 

\vspace{0.3cm}

\it Proof: \rm A simple induction takes care of the case $x_t\in\lbrace -\infty,\infty\rbrace$ for some $t$. We then note that due to the form of the recursion relation and the fact that the initial data is measurable and bounded, $G_t^i$ are differentiable for $t\geq 1$. Let us now assume that there is a point $x_t\in \R$ so that $G_t^2(x)> G_t^1(x)$ for $x> x_t$ and $G_t^2(x)< G_t^1(x)$ for $x<x_t$. Continuity implies that either there is a point $a\in \R$ so that  $G_{t+1}^2(a)=G_{t+1}^1(a)$ or $x_{t+1}\in\lbrace -\infty,\infty\rbrace$ so let us assume that such a point $a$ does exist. Our goal is to show that $x_{t+1}=a$ satisfies the desired conditions. Integrating by parts, 

\begin{equation*}
{G_{t+1}^{2}} '(a)-{G_{t+1}^{1}} '(a)=\int \rho(y) y \left(G_t^2(a+y)^K-G_t^1(a+y)^K\right)dy.
\end{equation*}

\vspace{0.3cm}

\noindent If $x_t\leq a$, then 

\begin{align*}
{G_{t+1}^{2}} '(a)-{G_{t+1}^1}'(a)&\geq \int_{\R\setminus (x_t-a,0)}\rho(y) y \left(G_t^2(a+y)^K-G_t^1(a+y)^K\right)dy\\
&+(x_t-a)\int_{x_t-a}^0\rho(y) \left(G_t^2(a+y)^K-G_t^1(a+y)^K\right)dy.
\end{align*}

\vspace{0.3cm}

\noindent $G_{t+1}^2(a)=G_{t+1}^1(a)$ implies that

\begin{align*}
\int_{x_t-a}^0 \rho(y)\left(G_t^2(a+y)^K-G_t^1(a+y)^K\right)dy=-\int_{\R\setminus (x_t-a,0)}\rho(y) \left(G_t^2(a+y)^K-G_t^1(a+y)^K\right)dy.
\end{align*}

\vspace{0.3cm}

\noindent Thus

\begin{align*}
{G_{t+1}^{2} }'(a)-{G_{t+1}^1}'(a)\geq \int_{\R\setminus (x_t-a,0)}\rho(y) (y-(x_t-a)) \left(G_t^2(a+y)^K-G_t^1(a+y)^K\right)dy.
\end{align*}

\vspace{0.3cm}

\noindent Since $x_t\leq a$, the definition of $x_t$ implies that the integrand is positive in the integration region, so ${G_{t+1}^{2}} '(a)-{G_{t+1}^1}'(a)> 0$. Similar reasoning shows that this result holds for $x_t\geq a$ as well.

\vspace{0.3cm}

\noindent Thus by induction, if for any $t>0$ there is a point $x_t\in \R$ so that $G_t^2(x_t)=G_t^1(x_t)$, then $G_t^2-G_t^1$ is strictly increasing at this point. Now if there was a point $x>x_t$ so that $G_t^2(x)<G_t^1(x)$, continuity and differentiability would imply that for some $x^*\in(x_t,x)$, $G_t^2(x^*)-G_t^1(x^*)=0$ and $G_t^2-G_t^1$ would be decreasing at this point. According to our reasoning, this is impossible so $G_t^2(x)\geq G_t^1(x)$ for $x>x_t$. 

\vspace{0.3cm}

\noindent If there was a point $x>x_t$ so that $G_t^2(x)=G_t^1(x)$, $G_t^2-G_t^1$ would be strictly increasing at this point. This means that $G_t^2-G_t^1$ would be negative in some open interval $(x-\delta,x)$, with $\delta>0$. This is contrary to $G_t^2(x)\geq G_t^1(x)$ for all $x\geq x_t$. Thus actually $G_t^2(x)>G_t^1(x)$ for all $x>x_t$. 

\vspace{0.3cm}

\noindent Since $G_t^2-G_t^1$ is strictly increasing at $x_t$, $G_t^2(x)-G_t^1(x)< 0$ for $x\in(x_t-\delta,x_t)$ for some $\delta>0$. If there were some $x<x_t$ so that $G_t^2(x)-G_t^1(x)\geq 0$, our previous reasoning would imply that $G_t^2(y)-G_t^1(y)>0$ for $y>x$, which is contrary to $G_t^2(y)-G_t^1(y)< 0$ for $y\in(x_t-\delta,x_t)$. $_{_\square}$

\vspace{0.3cm}

\bf Remark 2.6. \rm Looking at the proof, it is easy to see that if we change $>$ to $\geq $ and $<$ to $\leq$, the lemma will still be true.

\vspace{0.3cm}

\noindent This concludes the discussion about our main tools for proving convergence.

\vspace{1cm}

\section{Convergence for $\beta<\sqrt{2\log K}$}

In this section, we shall first consider the existence and asymptotics of traveling waves in the $c>\sqrt{2\log K}$ case and then demonstrate convergence for the case where the initial data is asymptotically $1-Ce^{-\beta x}$ (e.g., $\exp(-e^{-\beta x})$), with $\beta<\sqrt{2\log K}$. Convergence will also imply uniqueness of the traveling waves (uniqueness up to a translation). Many of the proofs are discrete time versions of those in \cite{Harris}. We will begin by showing that for each $c>\sqrt{2\log K}$, there exists a traveling wave with speed $c$. 

\vspace{0.3cm}

\bf Lemma 3.1. \rm For each $c>\sqrt{2\log K}$, there exists a function $w:\R\rightarrow[0,1]$ so that $w$ is increasing, $w(-\infty)=0$, $w(\infty)=1$ and 

\begin{equation*}
w(x)=\int \rho(y) w(x+y+c)^Kdy.
\end{equation*}

\vspace{0.3cm}

\it Proof: \rm According to Lemma 2.2, the positive martingales $Z_\beta(t)=\sum_{k=1}^{K^t} e^{-\beta(X_k(t)+c(\beta)t)}$ are uniformly integrable for $\beta<\sqrt{2\log K}$. Since they are positive, they converge. Let us write $Z_\beta(\infty)$ for the limit. We also define

\begin{equation*}
M_\beta(t)=\mathbb{E}\left.\left(e^{-Z_\beta(\infty)}\right|\mathcal{F}_t\right).
\end{equation*} 

\vspace{0.3cm}

\noindent Since $0\leq e^{-Z_\beta(\infty)}\leq 1$, $M_\beta$ is a uniformly integrable martingale converging to $e^{-Z_\beta(\infty)}$ almost surely. Let

\begin{equation*}
w_\beta(x)=\mathbb{E}^x\left(e^{-Z_\beta(\infty)}\right)=\mathbb{E}^0\left(e^{-e^{-\beta x}Z_\beta(\infty)}\right).
\end{equation*}

\vspace{0.3cm}

\noindent One can show (see \cite{Neveu}) that $Z_\beta(\infty)\in(0,\infty)$ almost surely for $\beta<\sqrt{2\log K}$ so $w_\beta(-\infty)=0$, $w_\beta(\infty)=1$ and $w_\beta$ is increasing. Thus we only need to show that $w_\beta$ is a traveling wave with speed $c(\beta)$. Using the decomposition of $Z_\beta(t+s)$ and passing to the limit, we have $Z_\beta(\infty)=\sum_{k=1}^{K^t}e^{-\beta(X_k(t)+c(\beta)t)}Z_\beta^{(k)}(\infty,t)$. Since the $Z_\beta^{(k)}$ are independent and identically distributed, we see that

\begin{align*}
M_\beta(t)&=\mathbb{E}\left.\left(e^{-Z_\beta(\infty)}\right|\mathcal{F}_t\right)\\
&=\prod_{k=1}^{K^t}\mathbb{E}\left.\left( \exp\left(-e^{-\beta(X_k(t)+c(\beta)t)}Z_\beta(\infty)\right)\right|\mathcal{F}_t\right)\\
&=\prod_{k=1}^{K^t}w_\beta(X_k(t)+c(\beta)t).
\end{align*}

\vspace{0.3cm}

\noindent So $\prod_{k=1}^{K^t} w_\beta(X_k(t)+c(\beta)t)$ is a uniformly integrable martingale, which implies that

\begin{equation*}
w_\beta(x)=\mathbb{E}^x\left(\prod_{k=1}^{K^t} w_\beta(X_k(t)+c(\beta)t)\right)
\end{equation*}

\vspace{0.3cm}

\noindent for each $t$. Setting $t=1$, we have 

\begin{equation*}
w_\beta(x)=\int \rho(y)w_\beta(x+y+c(\beta))^Kdy,
\end{equation*}

\vspace{0.3cm}

\noindent i.e. $w_\beta$ is a traveling wave with speed $c(\beta)$. Now for each $c>\sqrt{2\log K}$, we can find a $\beta<\sqrt{2\log K}$ so that $c=c(\beta)$. $_{_\square}$

\vspace{0.3cm}

The asymptotic behavior of a traveling wave in both cases ($c>\sqrt{2\log K}$ and $c=\sqrt{2\log K}$ follows from a result by Durrett and Liggett \cite{DL}. 

\vspace{0.3cm}

\bf Lemma 3.2. \rm If $w$ is a traveling wave of speed $c=c(\beta)>\sqrt{2\log K}$,  $w(-\infty)=0$ and $w(\infty)=1$, then $\lim_{x\rightarrow\infty}e^{\beta x}(1-w(x))=C$ for some $C>0$. 

\vspace{0.3cm}

For the proof, see \cite{DL}, Theorem 2.18 a). For the reader interested in going over the proof, we shall provide a short dictionary of what the different quantities appearing in \cite{DL} look like in our case. First of all, $W_1=W_2=...=W_K=e^{-\beta(V+c(\beta))}$ for all $i$, where $V$ is a standard Gaussian. Our solution to the stationary equation corresponds to their fixed point of a smoothing transformation via $w(x)=\phi(e^{-\beta x})$. Moreover, in our case the function $v(\alpha)$ is given by $v(\alpha)=\log K-\frac{1}{2}c(\beta)^2+\frac{1}{2}(\alpha\beta-c(\beta))^2$. Thus we see that for $\beta<\sqrt{2\log K}$, we are in the case where $v'(\alpha)<0$ if $v(\alpha)=0$ and for $\beta\geq \sqrt{2\log K}$ we are in the case $v'(\alpha)=0$ when $v(\alpha)=0$. Finally, the associated random walks are quite simple in our case. The increments of the walks are Gaussian and for $\beta<\sqrt{2\log K}$, the increments have positive expectation while for $\beta\geq \sqrt{2\log K}$, they are centered.

\vspace{0.3cm}

\vspace{0.3cm}

\noindent We note that if $w$ is a traveling wave, then $w(\ \cdot\ +a)$ is another traveling wave for each $a$. Thus the previous lemma implies that for each $\beta<\sqrt{2\log K}$ and each $C'>0$ we can find a traveling wave of speed $c(\beta)$ so that $e^{\beta x}(1-w(x))\rightarrow C'$ as $x\rightarrow \infty$. 

\vspace{0.3cm}

\noindent Our proof of convergence follows that of Lui's \cite{Lui}. Following Bramson's arguments, a complete classification of initial data for which convergence occurs is possible, but it requires more work. Before we demonstrate convergence, we need the following lemma. 

\vspace{0.3cm}

\bf Lemma 3.3. \rm Let $G_0(x)\geq \tilde{G}_0(x)-Ae^{-\beta x}$ for some $A\geq 0$ and $\beta>0$. Then for any $c>\sqrt{2\log K}$

\begin{equation*}
G_n(x+nc)\geq \tilde{G}_n(x+nc)-Ae^{-\beta x+n\beta\left(c(\beta)-c\right)}.
\end{equation*}

\vspace{0.3cm}

\it Proof: \rm The proof is by induction. We have

\begin{align*}
\tilde{G}_{n+1}(x+(n+1)c)-{G}_{n+1}(x+(n+1)c)&=\int \rho(x+c-y)\left(\tilde{G}_n(y+nc)^K-{G}_n(y+nc)^K\right)dy\\
&\leq K Ae^{n\beta\left(c(\beta)-c\right)}\int \rho(x+c-y)e^{-\beta y}dy\\
&=Ae^{(n+1)\beta\left(c(\beta)-c\right)}e^{-\beta x}._{_\square}
\end{align*}

\vspace{0.3cm}

\bf Theorem 3.4. \rm Let $e^{\beta x}(1-G_0(x))\rightarrow C>0$ as $x\rightarrow \infty$ and let $w$ be traveling wave of speed $c(\beta)$ satisfying $e^{\beta x}(1-w(x))\rightarrow C$ as $x\rightarrow\infty$ and $w(-\infty)=0$. Then $G_n(x+c(\beta)n)\rightarrow w(x)$ uniformly on sets of the form $[a,\infty)$ for any $a\in \R$. Moreover, if $G_0$ is increasing, then the convergence is uniform on $\R$. 

\vspace{0.3cm}

\it Proof: \rm For any $\delta>0$, we have

\begin{equation*}
\lim_{x\rightarrow\infty}\frac{1-G_0(x)}{1-w(x\pm \delta)}=e^{\pm\beta\delta}\lim_{x\rightarrow\infty}\frac{e^{\beta x}(1-G_0(x))}{e^{\beta (x\pm\delta)}(1-w(x\pm\delta))}=e^{\pm\beta\delta}.
\end{equation*}

\vspace{0.3cm}

\noindent Since $e^{-\beta \delta}<1<e^{\beta\delta}$, we can find a number $L_\delta$ so that $1-w(x+\delta)\leq 1-G_0(x)\leq 1-w(x-\delta)$, i.e., $w(x-\delta)\leq G_0(x)\leq w(x+\delta)$, for $x\geq L_\delta$. For any $b>0$, set $A_{b}=\sup_{x\leq L_\delta} e^{bx}(1-G_0(x))$ and $B_b=\sup_{x\leq L_\delta} e^{bx}(1-w(x+\delta))$. These definitions imply that

\begin{equation*}
w(x+\delta)\geq G_0(x)-B_b e^{-bx}
\end{equation*}

\vspace{0.3cm}

\noindent and

\begin{equation*}
G_0(x)\geq w(x-\delta)-A_be^{-bx}.
\end{equation*}

\vspace{0.3cm}

\noindent Using the previous lemma and the fact that if $\tilde{G}_0(x)=w(x)$ then $\tilde{G}_n(x+nc)=w(x)$, we have

\begin{equation*}
w(x-\delta)-A_be^{-bx+nb(c(b)-c(\beta))}\leq G_n(x+nc(\beta))\leq w(x+\delta)+B_be^{-bx+nb(c(b)-c(\beta))}.
\end{equation*}

\vspace{0.3cm}

\noindent Let $\epsilon>0$. Choosing $b\in(\beta,\sqrt{2\log K})$ implies $c(b)<c(\beta)$. Thus for any $a\in \R$, we can take $n$ so large that $\max\lbrace B_be^{-bx+nb(c(b)-c(\beta))},A_be^{-bx+nb(c(b)-c(\beta)}\rbrace<\frac{\epsilon}{2}$ for $x\geq a$. So we see that for $x\geq a$ and large enough $n$, 

\begin{equation*}
|w(x)-G_n(x+c(\beta)n)|\leq \frac{\epsilon}{2}+\max(w(x)-w(x-\delta),w(x+\delta)-w(x)).
\end{equation*}

\vspace{0.3cm}

\noindent We note that the equation of $w$ implies that $w$ is smooth and that $|w'(x)|<1$. Thus $w$ is uniformly continuous and we can take $\delta$ so small that $\max(w(x)-w(x-\delta),w(x+\delta)-w(x))\leq \frac{\epsilon}{2}$ and 

\begin{equation*}
|w(x)-G_n(x+c(\beta)n)|\leq \epsilon
\end{equation*}

\vspace{0.3cm}

\noindent for $x\geq a$ and large enough $n$. So we see that $G_n(x+c(\beta)n)\rightarrow w(x)$ uniformly on sets of the form $[a,\infty)$. If $G_0$ is increasing, then $G_n$ is increasing for each $n$. Let $\epsilon>0$ and set $a$ so that $w(x)\leq \frac{\epsilon}{2}$ for $x\leq a$. We then take $n$ so large that $|G_n(x+nc(\beta))-w(x)|<\frac{\epsilon}{2}$ for $x\geq a$. Thus $G_n(a+c(\beta)n)\leq \epsilon$. Since $G_n$ is increasing, $G_n(x+nc(\beta))\leq \epsilon$ for $x\leq a$. Since $G_n$ and $w$ are non-negative, we see that $|G_n(x+nc(\beta))-w(x)|\leq\epsilon$ for $x\leq a$ and we have uniform convergence on $\R$.$_{_\square}$

\vspace{0.3cm}

\bf Corollary 3.5. \rm For each $C>0$ there is only one traveling wave of speed $c>\sqrt{2\log K}$ so that $w(-\infty)=0$ and $e^{\beta x}(1-w(x))\rightarrow C$ as $x\rightarrow\infty$. Moreover, this traveling wave is increasing.

\vspace{0.3cm}

\it Proof: \rm In the proof of the previous result, we did not fix which traveling wave $w$ we are using apart from fixing $C$. Since any such $w$ would then be the limit of the sequence $G_n(x+c(\beta)n)$, we see that there can be only one such $w$. Taking $G_0(x)=\exp(-e^{-\beta x})$, a simple induction shows that $G_t'(x)>0$ for all $x$ so we have a sequence of increasing functions converging to $w$ so the limit must be increasing.  $_{_\square}$

\vspace{0.3cm}

\noindent This concludes our discussion about the $\beta<\sqrt{2\log K}$ case. 

\vspace{1cm}

\section{Convergence for Heaviside initial data}

\vspace{0.3cm}

We will now focus on Heaviside initial data, i.e., the $\beta=\infty$ case. This will be important to us in the next section when we prove convergence in the $\beta\geq \sqrt{2\log K}$ case. To get started, we note that it is simple to check that $G_t^{-1}:(0,1)\rightarrow\R$ is well defined for $t\geq 1$. Thus we can define $m_y(t)=G_t^{-1}(y)$. 

\vspace{0.3cm}

\noindent Convergence of $G_t(x+m_y(t))$ follows from Lemma 2.5:

\vspace{0.3cm}

\bf Lemma 4.1. \rm Let $G_t$ be given by the recursion relation \eqref{eq:recursion} with Heaviside initial data and let $y\in(0,1)$ be fixed. Then the limit function $w_y(x)=\lim_{t\rightarrow\infty}G_t(x+m_{y}(t))$ exists. 

\vspace{0.3cm}

\it Proof: \rm Let $t_0\in\mathbb{N}\setminus\lbrace 0\rbrace$ be fixed. Let us set $G_t^1(x)=G_{t+1}(x+m_y(t_0+1))$ and $G_t^2(x)=G_t(x+m_y(t_0))$. Now $0<G_1(x)<1$ for all $x$ so we see that $G_0^2(x)<G_0^1(x)$ for $x<-m_y(t_0)$ and $G_0^2(x)>G_0^1(x)$ for $x>-m_y(t_0)$, so the conditions of Lemma 2.5 are met.

\vspace{0.3cm}

\noindent Moreover,

\begin{equation*}
G_{t_0}^2(0)=G_{t_0}(m_y(t_0))=y=G_{t_0+1}(m_y(t_0+1))=G_{t_0}^1(0).
\end{equation*}

\vspace{0.3cm}

\noindent As $t_0$ is arbitrary, Lemma 2.5 then implies that for $x>0$, $(G_t(x+m_y(t)))_t$ is decreasing and bounded from below by $0$ so there must be a limit which we call $w_y(x)$. In a similar manner, we see that $(G_{t}(m_y(t)+x))_t$ is increasing for $x<0$ and constant for $x=0$ so the limiting function $w_y:\R\rightarrow\R$ exists. $_{_\square}$

\vspace{0.3cm}

\noindent From now on we shall fix $y=\frac{1}{2}$ and write $m_\frac{1}{2}(t)=m(t)$ as well as $w=w_\frac{1}{2}$. We note that $w$ is increasing and $w(0)=\frac{1}{2}$. We shall also write $\Delta m(t)=m(t+1)-m(t)$. To prove that the limit $w$ is a traveling wave, we need some simple properties of $m$. For example, linearizing the recursion of $1-G_t$, one can check that there is a constant $C$ so that $m(t)\leq \sqrt{2\log K} t-2^{-\frac{3}{2}}\frac{\log t}{\sqrt{\log K}}+C$. Moreover, a simple argument using the form of the recursion relation and the fact that $G_t(x+m(t))$ increases to $w(x)$ for $x\leq 0$, implies that $(\Delta m(t))_t$ is bounded from below. This is in fact enough to show that $w$ is a traveling wave for some $c$.

\vspace{0.3cm}

\bf Lemma 4.2. \rm There is a unique $c$ so that the limit function $w$ is a traveling wave with speed $c$.

\vspace{0.3cm}

\it Proof: \rm We first note that we can find a subsequence of $(\Delta m(t))_t$ that converges to some finite value. Otherwise $m(t)\leq \sqrt{2\log K} t-2^{-\frac{3}{2}}\frac{\log t}{\sqrt{\log K}}+C$ would be violated, since we know $(\Delta m(t))_t$ to be bounded from below. Let the limit of this subsequence be $c$.

\vspace{0.3cm}

\noindent From the recursion relation, one can check that $0\leq G_t'(x)\leq 1$ for all $x$ and $t$. Thus we have

\begin{align*}
|G_{t_k+1}(x+m(t_k))-w(x-c)|\leq |\Delta m(t_k)-c|+|G_{t_k+1}(x-c+m(t_k+1))-w(x-c)|\rightarrow 0,
\end{align*}

\vspace{0.3cm}

\noindent as $k\rightarrow \infty$ for each fixed $x$. On the other hand

\begin{align*}
G_{t_k+1}(x+m(t_k))=\int \rho(y) G_{t_k}(x+y+m(t_k))^K dy\rightarrow \int \rho(y) w(x+y)^Kdy
\end{align*}

\vspace{0.3cm}

\noindent as $k\rightarrow\infty$. We conclude that $w$ satisfies

\begin{equation*}
w(x-c)=\int \rho(y) w(x+y)^Kdy,
\end{equation*}

\vspace{0.3cm}

\noindent which is precisely the equation we wanted. To show that the value $c$ does not depend on the subsequence we picked, let us assume that there were two such values $c$ and $c'$. Since $w$ is increasing, $G_1(x)>0$ for all $x$ and $G_t(x+m(t))$ increases to $w(x)$ for $x\leq 0$, $w(x)>0$ for all $x$. Also $w'(x)=\int \rho(y) K w(x+y+c)^{K-1}w'(x+y+c)dy$ so for a given $x$, $w'(x)$ can be zero only if $w$ is a constant. If $w$ were a constant function, we would have $w(x)=w(0)=\frac{1}{2}$, which does not satisfy the equation for $w$.  Thus $w$ is strictly increasing. Since

\begin{equation*}
w(x-c)=\int \rho(y) w(x+y)^Kdy=w(x-c').
\end{equation*}

\vspace{0.3cm}

\noindent and $w$ is strictly increasing, $c=c'$. $_{_\square}$

\vspace{0.3cm}

\noindent It follows from the recursion relation that $|G_t^{(n)}(x)|\leq \sqrt{n!}$ and $|w^{(n)}(x)|\leq \sqrt{n!}$ for all $n$, $x$ and $t\geq 1$, which implies that all of these functions are entire. Using some basic results from complex analysis, it then follows that the convergence to $w$ is uniform on $\R$. Using this uniform convergence and the recursion relation, one can argue that $w(x-\Delta m(t))\rightarrow w(x-c)$, which implies that $\Delta m(t)\rightarrow c$ and $c\leq \sqrt{2\log K}$.

\vspace{0.3cm}

\noindent To show that $c=\sqrt{2\log K}$, we will make use of the branching random walk.

\vspace{0.3cm}

\bf Lemma 4.3. \rm $W(t)=\prod_{k=1}^{K^t} w(X_k(t)+ct)$ is a martingale with respect to the branching random walk and $w(x)=\mathbb{E}^x(W(t))$ for all $t$.  

\vspace{0.3cm}

\it Proof: \rm Decomposing the product into the product over the clusters and the product over particles inside each cluster and using independence we have

\begin{align*}
\mathbb{E}^x(W(t+1)|\mathcal{F}_t)&=\prod_{k=1}^{K^t} \int \rho(y)w(X_k(t)+ct+y+c)^Kdy\\
&=\prod_{k=1}^{K^t} w(X_k(t)+ct)\\
&=W(t).
\end{align*}

\vspace{0.3cm}

\noindent Since $W$ is a martingale, $\mathbb{E}^x(W(t))=\mathbb{E}^x(W(0))=w(x)$. $_{_\square}$

\vspace{0.3cm}

\bf Lemma 4.4. \rm $c=\sqrt{2\log K}$.

\vspace{0.3cm}

\it Proof: \rm Let us assume that $c<\sqrt{2\log K}$. The martingale $W$ that we introduced in the previous lemma is positive so it converges. Let us denote the limit by $W(\infty)$. Moreover, it is bounded above by one so it is also uniformly integrable and $w(x)=\mathbb{E}^x(W(\infty))$. On the other hand

\begin{equation*}
0\leq W(t)=\prod_{k=1}^{K^t} w(X_k(t)+ct)\leq w(L(t)+ct).
\end{equation*}

\vspace{0.3cm}

\noindent According to Lemma 2.3, $L(t)+ct\rightarrow -\infty$ almost surely as $t\rightarrow\infty$. Since $w(x)\rightarrow 0$ as $x\rightarrow-\infty$, this implies that $W(\infty)=0$ almost surely and $w(x)=\mathbb{E}^x(W(\infty))=0$ for all $x$, which is contrary to our knowledge of $w$ being a function increasing from zero to one. Thus $c=\sqrt{2\log K}$. $_{_\square}$

\vspace{0.3cm}

\noindent For the asymptotic behavior of $w$, we rely on \cite{DL} again (Theorem 2.18 b)) as in the case of Lemma 3.2.

\vspace{0.3cm}

\bf Lemma 4.5. \rm Let $w$ be a traveling wave with speed $c=\sqrt{2\log K}$. Then \rm $\frac{e^{\sqrt{2\log K}x}(1-w(x))}{x}\rightarrow C$ as $x\rightarrow\infty$ for some $C>0$.

\vspace{0.3cm}

\noindent This along with our discussion about the branching random walk gives the uniqueness of the traveling waves.

\vspace{0.3cm}

\bf Lemma 4.6. \rm Every traveling wave $\tilde{w}$ with speed $c=\sqrt{2\log K}$ satisfying $\tilde{w}(-\infty)=0$ and $\tilde{w}(\infty)=1$, is given by a translation of the limit of the Heaviside case $w$.

\vspace{0.3cm}

\it Proof: \rm Let $\tilde{w}:\R\rightarrow (0,1)$ be any non-trivial solution to the equation

\begin{equation*}
\tilde{w}(x)=\int \rho(y) \tilde{w}(x+y+c)^Kdy
\end{equation*}

\vspace{0.3cm}

\noindent with $\tilde{w}(-\infty)=0$ and $\tilde{w}(\infty)=1$. Also let 

\begin{equation*}
W^y(t)=\prod_{k=1}^{K^t} \tilde{w}(X_k(t)+ct+y).
\end{equation*}

\vspace{0.3cm}

\noindent As in Lemma 4.3, one can show that $W^y$ is a uniformly integrable martingale which converges (to say $W^y(\infty)$). By Lemma 4.5 there is a $\tilde{x}\in \R$ so that $\tilde{w}(x)\sim 1-xe^{-c(x+\tilde{x})}$. Thus for each fixed $y$ we have

\begin{align*}
\lim_{x\rightarrow\infty}\frac{xe^{-c(x+y+\tilde{x})}}{-\log \tilde{w}(x+y)}&=1,
\end{align*}

\vspace{0.3cm}

\noindent i.e. for each fixed $y\in\R$ and $\epsilon>0$ we can find a $D\in \R$ so that for $x\geq D$

\begin{equation*}
1-\epsilon\leq \frac{xe^{-c(x+y+\tilde{x})}}{-\log \tilde{w}(x+y)}\leq 1+\epsilon.
\end{equation*}

\vspace{0.3cm}

\noindent Using Lemma 2.4, we see that taking $t$ large enough, $L(t)+ct\geq D$ almost surely. This means that these inequalities hold when we set $x$ to be $X_k(t)+ct$ for any $k$. Then summing over all $k$ we obtain

\begin{align*}
(1-\epsilon)(-\log(W^y(t)))&\leq e^{-c(y+\tilde{x})}\sum_{k=1}^{K^t}(X_k(t)+ct)e^{-c(X_k(t)+ct)}\\
&\leq (1+\epsilon)(-\log W^y(t)).
\end{align*}

\vspace{0.3cm}

\noindent This means that

\begin{equation*}
W^y(\infty)=\exp\left(e^{-c(y+\tilde{x})}\lim_{t\rightarrow\infty}\left.\frac{\partial}{\partial \beta}Z_\beta(t)\right|_{\beta=c}\right)
\end{equation*}

\vspace{0.3cm}

\noindent Since $W^y$ is uniformly integrable, we have

\begin{equation*}
\tilde{w}(y)=\mathbb{E}^0(W^y(\infty))=\mathbb{E}^0\left(\exp\left(e^{-c(y+\tilde{x})}\lim_{t\rightarrow\infty}\left.\frac{\partial}{\partial \beta}Z_\beta(t)\right|_{\beta=c}\right)\right).
\end{equation*}

\vspace{0.3cm}

\noindent Hence

\begin{equation*}
\tilde{w}(y-\tilde{x})=\mathbb{E}^0\left(\exp\left(e^{-cy}\lim_{t\rightarrow\infty}\left.\frac{\partial}{\partial \beta}Z_\beta(t)\right|_{\beta=c}\right)\right).
\end{equation*}

\vspace{0.3cm}

\noindent The right side of this equation is completely independent of the solution of the stationary equation we pick. Thus every solution must be a translation of $w$. $_{_\square}$

\vspace{0.3cm}

\noindent This concludes our treatment of the Heaviside case.

\vspace{1cm}

\section{Convergence for $\beta\geq\sqrt{2\log K}$}

\vspace{0.3cm}

\noindent In this section, we shall demonstrate convergence for initial data $G_0(x)=\exp(-e^{-\beta x})$ with $\beta\geq \sqrt{2\log K}$. Our main tool will be Lemma 2.5 along with the knowledge of convergence in the cases $\beta<\sqrt{2\log K}$ and $\beta=\infty$. 

\vspace{0.3cm}

\noindent Let us write $G_t^\beta$ for the solution of the recursion relation with initial data $G_0^\beta(x)=\exp(-e^{-\beta x})$. One can check that $m^\beta(t)=(G_t^{\beta})^{-1}(\frac{1}{2})$ is well defined. 

\vspace{0.3cm}

\bf Lemma 5.1. \rm For $\beta>\beta'$, $G_t^\beta(x+m^\beta(t))\geq G_t^{\beta'}(x+m^{\beta'}(t))$ for $x\geq 0$ and $G_t^\beta(x+m^\beta(t))\leq G_t^{\beta'}(x+m^{\beta'}(t))$ for $x\leq 0$. 

\vspace{0.3cm}

\it Proof: \rm Let us fix $t_0$ and set $\tilde{G}_t^\beta(x)=G_t^\beta(x+m^\beta(t_0))$ and $\tilde{G}_t^{\beta'}(x)=G_t^{\beta'}(x+m^{\beta'}(t_0))$. Now $\tilde{G}_0^\beta(x)\geq \tilde{G}_0^{\beta'}(x)$ if 

\begin{equation*}
x\geq \frac{\beta' m^{\beta'}(t_0)-\beta m^\beta(t_0)}{\beta-\beta'}
\end{equation*}

\vspace{0.3cm}

\noindent and $\tilde{G}_0^\beta(x)\leq \tilde{G}_0^{\beta'}(x)$ if $x\leq \frac{\beta' m^{\beta'}(t_0)-\beta m^\beta(t_0)}{\beta-\beta'}$. So by Lemma 2.5, there is a $x_t$ so that $\tilde{G}_t^\beta(x)\geq\tilde{G}_t^{\beta'}(x)$ for $x\geq x_t$ and $\tilde{G}_t^\beta(x)\leq\tilde{G}_t^{\beta'}(x)$ for $x< x_t$. Since $\tilde{G}_{t_0}^\beta(0)=\frac{1}{2}=\tilde{G}_{t_0}^{\beta'}(0)$, we see that $x_{t_0}=0$. Since $t_0$ was arbitrary, this proves the lemma. $_{_\square}$ 

\vspace{0.3cm}

\bf Corollary 5.2. \rm For any $\epsilon>0$ and large enough $t$, $G_t^\beta(x+m^\beta(t))\leq w(x)+\epsilon$ for $x\geq 0$ and $G_t^\beta(x+m^\beta(t))\geq w(x)-\epsilon$ for $x\leq 0$, where $w$ is the limit of the Heaviside case.

\vspace{0.3cm}

\it Proof: \rm This follows by setting $\beta=\infty$ in the previous lemma and then using the uniform convergence of the Heaviside case.$_{_\square}$

\vspace{0.3cm}

\noindent From now on, we shall write $w_c$ for a traveling wave with speed $c$ and we shall write $w$ for the limit in the Heaviside case.

\vspace{0.3cm}

\bf Lemma 5.3. \rm For $\beta<\sqrt{2\log K}$, $G_t^\beta(x+m^\beta(t))$ converges to the traveling wave $w_{c(\beta)}$ with speed $c(\beta)$ and normalized to $w_{c(\beta)}(0)=\frac{1}{2}$.

\vspace{0.3cm}

\it{Proof: } \rm We know that $G_t^\beta(x+c(\beta)t)$ converges to a traveling wave $\tilde{w}_{c(\beta)}$ with speed $c(\beta)$ uniformly. Thus $G_t^\beta(x+m^\beta(t))=\tilde{w}_{c(\beta)}(x+m^\beta(t)-c(\beta)t)+\mathit{o}(1)$. Setting $x=0$, we have $\frac{1}{2}=\tilde{w}_{c(\beta)}(m^\beta(t)-c(\beta)t)+\mathit{o}(1)$. Passing to the limit we see that the limit $\lim_{t\rightarrow \infty}(m(t)-c(\beta)t)$ exists. This implies that

\begin{equation*}
\lim_{t\rightarrow\infty}G_t^\beta(x+m^\beta(t))=\tilde{w}_{c(\beta)}(x+\alpha),
\end{equation*}

\vspace{0.3cm}

\noindent and $\alpha$ is determined by the condition $\tilde{w}_{c(\beta)}(\alpha)=\frac{1}{2}$. Now of course $w_{c(\beta)}(x):=\tilde{w}_{c(\beta)}(x+\alpha)$ is a traveling wave as well. $_{_\square}$

\vspace{0.3cm}

\noindent We are now ready to prove convergence for $\beta\geq \sqrt{2\log K}$.

\vspace{0.3cm}

\bf Theorem 5.4. \rm For $\beta\geq \sqrt{2\log K}$, $G_t^\beta(x+m^\beta(t))$ converges uniformly to $w(x)$, where $w$ is the limit of the Heaviside case.

\vspace{0.3cm}

\it Proof: \rm Lemma 5.1 implies that for $\beta_1<\beta_2<\sqrt{2\log K}$ we have ${w}_{c(\beta_1)}(x)\leq {w}_{c(\beta_2)}(x)$ for $x\geq 0$ and ${w}_{c(\beta_1)}(x)\geq {w}_{c(\beta_2)}(x)$ for $x\leq 0$. Also ${w}_{c(\beta_i)}(x)\leq w(x)$ for $x\geq 0$ and ${w}_{c(\beta_i)}(x)\geq w(x)$ for $x\leq 0$. This implies that we have a pointwise limit 

\begin{equation*}
\hat{w}(x)=\lim_{c\rightarrow \sqrt{2\log K}\ ^{+}}{w}_c(x).
\end{equation*}

\vspace{0.3cm}

\noindent Since $w_c$ are increasing functions, $\hat{w}$ is increasing as well. Now we have for small $\delta>0$

\begin{align*}
{w}_{\sqrt{2\log K}+\delta}(x-\sqrt{2\log K}-\delta)=\int \rho(y){w}_{\sqrt{2\log K}+\delta}(x+y)^Kdy.
\end{align*}

\vspace{0.3cm}

\noindent Since $0\leq {w}_{c}'(x)\leq 1$, we see that

\begin{equation*}
{w}_{\sqrt{2\log K}+\delta}(x-\sqrt{2\log K})+\mathcal{O}(\delta)=\int \rho(y){w}_{\sqrt{2\log K}+\delta}(x+y)^Kdy.
\end{equation*}

\vspace{0.3cm}

\noindent Taking the limit $\delta\rightarrow 0$, we see that

\begin{equation*}
\hat{w}(x-\sqrt{2\log K})=\int \rho(y)\hat{w}(x+y)^Kdy.
\end{equation*}

\vspace{0.3cm}

\noindent Since up to translation, this equation has a unique increasing solution and $\hat{w}(0)=\frac{1}{2}$, we see that $\hat{w}=w$. Now Lemma 5.1 implies that we have the pointwise estimate ${w}_{\sqrt{2\log K}+\delta}(x)-\epsilon\leq G_t^\beta(x+m^\beta(t))\leq w(x)+\epsilon$ for $x\geq 0$, $\beta\geq \sqrt{2\log K}$, $\delta>0$ and large enough $t$. Then taking $\delta\rightarrow 0$ we see that pointwise $G_t^\beta(x+m^\beta(t))\rightarrow w(x)$ for $x\geq 0$. We see this in a similar manner for $x\leq 0$.

\vspace{0.3cm}

\noindent We can actually extend this pointwise estimate to a uniform one. Let us consider the sequence $f_n(x)=w_{\sqrt{2\log K}+\frac{1}{n}}(x)$. We know that pointwise $f_n(x)$ increases to $w(x)$ for $x\geq 0$ and decreases to it for $x\leq 0$. Also we know that $f_n(x)\rightarrow 1$ as $x\rightarrow \infty$ and $f_n(x)\rightarrow 0$ as $x\rightarrow -\infty$. Let $\epsilon>0$. We can take $N$ so large that 

\begin{equation*}
0<1-w(x)\leq 1-f_n(x)\leq 1-f_1(x)<\epsilon
\end{equation*}

\vspace{0.3cm}

\noindent for $x\geq N$ and 

\begin{equation*}
0<w(x)\leq f_n(x)\leq f_1(x)<\epsilon
\end{equation*}

\vspace{0.3cm}

\noindent for $x\leq -N$. Now $0<f_n'(x)<1$ and $0<f_n(x)<1$ for all $x$. This implies that the sequence $(f_n)$ is uniformly bounded and equicontinuous. Thus we can pick a subsequence $(f_{n_k})_k$ that converges uniformly to $w$ on compact sets. We thus have for large enough $t$ and $k$

\begin{equation*}
w_{\sqrt{2\log K}+\frac{1}{n_k}}(x)-\epsilon\leq G_t^\beta(x+m^\beta(t))\leq w(x)+\epsilon.
\end{equation*}

\vspace{0.3cm}

\noindent uniformly on $[0,\infty)$. Taking $k\rightarrow \infty$ we see that $G_t(x+m(t))\rightarrow w(x)$ uniformly on $[0,\infty)$. The proof for $(-\infty,0]$ is similar. $_{_\square}$

\vspace{0.3cm}

Knowing convergence to the traveling wave allows us to extract the leading order contribution to $m^\beta(t)$.

\vspace{0.3cm}

\bf Lemma 5.5. \rm If $G_t(x+f(t))\rightarrow w(x)$ uniformly for some $f(t)$, where $w$ is a traveling wave with speed $c\geq\sqrt{2\log K}$, then $f(t+s)-f(t)\rightarrow cs$ for any fixed $s$ and $\frac{f(t)}{t}\rightarrow c$.

\vspace{0.3cm}

\it Proof: \rm First of all we note that a simple indeuction implies that for any two initial data (measurable and between 0 and 1) $G_0$ and $\tilde{G}_0$, $\sup_{x}|G_t(x)-\tilde{G}_t(x)|\leq K^t \sup_x |G_0(x)-\tilde{G}_0(x)|$. Also $\tilde{G}_t(x)=w(x-ct)$ satisfies \eqref{eq:recursion}. Combining these two remarks, we see that for any fixed $s$, $\sup_x|G_{t+s}(x+f(t)+cs)-w(x)|\leq K^s\sup_x |G_t(x+f(t))-w(x)|\rightarrow 0$. On the other hand, also $G_{t+s}(x+f(t+s))\rightarrow w(x)$ uniformly for each $s$, so comparing the two sequences of functions one can argue that $f(t+s)-f(t)\rightarrow cs$ for each $s$. This in turn implies that $\frac{f(t)}{t}\rightarrow c$.$_{_\square}$

\section{The discrete time Brownian bridge and lower order terms for $m^\beta(t)$}

\vspace{0.3cm}

Up to now, we have showed that $G_t(x+m^\beta(t))$ converges to a traveling wave uniformly. We have also showed that for $\beta<\sqrt{2\log K}$, $m^\beta(t)-c(\beta)t\rightarrow C$ for some constant $C$. We know that for $\beta\geq \sqrt{2\log K}$, the leading order term in $m^\beta(t)$ is $\sqrt{2\log K}t$. So our next goal is to find the lower order terms of $m^\beta(t)$ for $\beta\geq \sqrt{2\log K}$.

\vspace{0.3cm}

As mentioned in the introduction, Bramson has done this in the continuum time case. In discrete time, the problem can be solved with very similar arguments. We shall not repeat all of his arguments, but merely formulate the problem in discrete time in a similar manner as the continuum problem, prove an estimate that is very important in many other estimates and then give a brief sketch of the argument.

\vspace{0.3cm}

Bramson's main tool in analyzing the KPP-equation is the Feynman-Kac formula. He uses this to represent the solution of the equation in terms of an expectation with respect to the Brownian bridge. To derive the Feynman-Kac formula in discrete time, let us write $U_t=1-G_t$ and iterate the recursion relation for $U$. One obtains

\begin{align*}
U_t(x)&=\int_{\R^t} \left(\prod_{s=1}^t \rho(y_s)\right)U_0\left(x+\sum_{s=1}^t y_s\right)e^{\sum_{s=1}^t k_{t-s}\left(x+\sum_{i=1}^s y_i\right)}\prod_{s=1}^t dy_s,
\end{align*} 

\vspace{0.3cm}

\noindent where $k_s(y)=\log\sum_{j=0}^{K-1} G_s(y)^j$. We can of course interpret $X_t=x+\sum_{s=1}^t y_s$ as a random walk (actually discrete time Brownian motion). Let us write $\mathbf{P}^x$ for the law of this random walk starting at $x$ and $\mathbf{E}^x$ for the expectation with respect to it. So we see that

\begin{equation*}
U_t(x)=\mathbf{E}^x\left(U_0(X_t)e^{\sum_{s=1}^t k_{t-s}(X_s)}\right).
\end{equation*}

\vspace{0.3cm}

\noindent We can then split the expectation so that we consider random walks from $x$ to $y$ and average over the end point $y$ so we have

\begin{equation}\label{eq:feynmankac}
U_t(x)=\int_{\R} \frac{1}{\sqrt{2\pi t}}e^{-\frac{(x-y)^2}{2t}}U_0(y)\mathbb{E}_t^{x,y}\left(e^{\sum_{s=1}^t k_{t-s}(Y_s)}\right)dy,
\end{equation}

\vspace{0.3cm}

\noindent where $Y$ is a random walk from $x$ to $y$ in $t$ steps with normalized Gaussian increments. $\mathbb{E}_t^{x,y}$ is the expectation with respect to this random walk and we shall also write $\mathbb{P}_t^{x,y}$ for the law of it.

\vspace{0.3cm}

\noindent One can check that the density of the joint distribution of $(Y_{k_1},...,Y_{k_s})$ (where $0<k_i<k_{i+1}<t$) is 

\begin{equation}\label{eq:density}
p_{ k_1,...,k_s}^{x,y}(y_{1},...,y_{s})=\sqrt{\frac{t}{t-k_s}}e^{\frac{(x-y)^2}{2t}} \prod_{j=1}^{s+1}\frac{1}{\sqrt{2\pi (k_j-k_{j-1})}}e^{-\frac{\left(y_{j}-y_{{j-1}}\right)^2}{2(k_{j}-k_{j-1})}},
\end{equation}

\vspace{0.3cm}

\noindent where $k_0=0$, $k_{s+1}=t$, $y_0=x$ and $y_{s+1}=y$. Another way to view this process is that it is a continuum time Brownian bridge evaluated at integer times. Because of this interpretation, we shall write $\mathbb{P}_t^{x,y}$ for the law of the continuum time Brownian bridge as well and indicate whether we are interested in the continuum or discrete time in some other way.

\vspace{0.3cm}

In analyzing the recursion relation through the Feynman-Kac formula, the main task is to determine which paths contribute significantly to the expectation. In the continuum time case, Bramson does this by calculating probabilities for the Brownian bridge to hit different sets. This involves quite a few technical details and a fair amount of work. In the end, almost all of these estimates rely on the fact that one can calculate the following probability exactly 

\begin{equation}\label{eq:exact}
\mathbb{P}_t^{x,y}(Y(s)>0 \ \mathrm{for} \ s\in[0,t])=1-e^{-\frac{2xy}{t}}.
\end{equation}

\vspace{0.3cm}

\noindent The proof of this can be found for example in Bramson's work. As mentioned in the introduction, the analysis in the discrete time case is formally identical to the continuum time case and we shall not be going over the technical details here. What we will do is to demonstrate that one can use $\mathbb{P}_t^{x,y}(Y(s)>0 \ \mathrm{for} \ s\in\lbrace 0,...,t\rbrace)$ as in the continuum case. To do this, we shall need the following result that Bramson proves.

\vspace{0.3cm}

\bf Lemma 6.1. \rm Let $l_1,l_2:[0,t]\rightarrow [-\infty,\infty]$ be upper semi-continuous functions satisfying $l_1(s)\leq l_2(s)$ for all $s\in [0,t]$. Then

\begin{equation*}
\frac{{\mathbb{P}}_t^{x,y}(Y(s)>l_2(s) \ \mathrm{for} \ s\in[0,t])}{{\mathbb{P}}_t^{x,y}(Y(s)>l_1(s) \ \mathrm{for} \ s\in[0,t])}
\end{equation*}

\vspace{0.3cm}

\noindent is increasing in $x$ and $y$.

\vspace{0.3cm}

\noindent We note that this result also contains the discrete time case since if $l$ is defined on $\lbrace 0,...,t\rbrace$, it can be extended to an upper semi-continuous function on $[0,t]$ by setting $l(s)=-\infty$ for non-integer values of $s$. 

\vspace{0.3cm}

Using this lemma, we can prove the required estimate in the discrete time case. The approach to the proof was suggested by Greg Lawler. For brevity, let us write $\mathbb{P}_t^{x,y}(B_0)$ for $\mathbb{P}_t^{x,y}(Y(s)>0 \ \mathrm{for} \ s\in\lbrace 0,...,t\rbrace)$ in this proof.

\vspace{0.3cm}

\bf Lemma 6.2. \rm There exists a constant $C>0$ so that for $x,y\geq 0$,

\begin{equation}\label{eq:ineq}
\mathbb{P}_t^{x,y}(B_0)\leq C\frac{(1+x)(1+y)}{t}.
\end{equation} 

\vspace{0.3cm}

\noindent There is also a constant $C'>0$ so that if $x,y\geq 0$ and $xy\leq t$, then 

\begin{equation*}
\mathbb{P}_t^{x,y}(B_0)\geq C'\frac{xy}{t}.
\end{equation*}

\vspace{0.3cm}

\it Proof: \rm  The lower bound is just an elementary estimate related to the corresponding continuum quantity \eqref{eq:exact}: we have 

\begin{equation*}
\mathbb{P}_t^{x,y}(B_0)\geq \mathbb{P}_t^{x,y}(Y(s)>0 \ \mathrm{for} \ s\in[0,t])=1-e^{-2\frac{xy}{t}}.
\end{equation*}

\vspace{0.3cm}

\noindent For the upper bound, we split the random walk into three parts of length $\frac{t}{3}$:

\begin{align*}
\mathbb{P}_t^{x,y}(B_0)&= \sqrt{2\pi t}e^{\frac{(x-y)^2}{2t}}\int_0^\infty\int_0^\infty  \frac{1}{\sqrt{2\pi\lfloor\frac{t}{3}\rfloor}}e^{-\frac{(z_1-x)^2}{2\lfloor \frac{t}{3}\rfloor}}\mathbb{P}_{\lfloor\frac{t}{3}\rfloor}^{x,z_1}(B_0)\\
&\times\frac{1}{\sqrt{2\pi\left(t-2\lfloor\frac{t}{3}\rfloor\right)}}e^{-\frac{(z_2-z_1)^2}{2\left(t-2\lfloor \frac{t}{3}\rfloor\right)}}\mathbb{P}_{t-2\lfloor\frac{t}{3}\rfloor}^{x,z_1}(B_0)\frac{1}{\sqrt{2\pi\lfloor\frac{t}{3}\rfloor}}e^{-\frac{(z_2-y)^2}{2\lfloor \frac{t}{3}\rfloor}}\mathbb{P}_{\lfloor\frac{t}{3}\rfloor}^{z_2,y}(B_0)dz_1dz_2\\
&\leq \sqrt{2\pi t}e^{\frac{(x-y)^2}{2t}}\int_0^\infty\int_0^\infty  \frac{1}{\sqrt{2\pi\lfloor\frac{t}{3}\rfloor}}e^{-\frac{(z_1-x)^2}{2\lfloor \frac{t}{3}\rfloor}}\mathbb{P}_{\lfloor\frac{t}{3}\rfloor}^{x,z_1}(B_0)\frac{1}{\sqrt{2\pi\left(t-2\lfloor\frac{t}{3}\rfloor\right)}}\\
&\times\frac{1}{\sqrt{2\pi\lfloor\frac{t}{3}\rfloor}}e^{-\frac{(z_2-y)^2}{2\lfloor \frac{t}{3}\rfloor}}\mathbb{P}_{\lfloor\frac{t}{3}\rfloor}^{z_2,y}(B_0)dz_1dz_2\\
&\leq C e^{\frac{(x-y)^2}{2t}}\mathbf{P}^x\left(X(s)>0 \ \mathrm{for} \ s\in\left\lbrace 0,...,\left\lfloor\frac{t}{3}\right\rfloor\right\rbrace\right)\mathbf{P}^y\left(X(s)>0 \ \mathrm{for} \ \left\lbrace 0,...,\left\lfloor\frac{t}{3}\right\rfloor\right\rbrace\right).
\end{align*}

\vspace{0.3cm}

\noindent By the Gambler's ruin estimate (see \cite{Lawler}), there is a constant $\tilde{C}>0$ so that for $x\leq \sqrt{t}$, 

\begin{equation*}
\mathbf{P}^x(X(s)>0 \ \mathrm{for}\ s\in\lbrace 0,...,t\rbrace)\leq \tilde{C}\frac{1+x}{\sqrt{t}}.
\end{equation*}

\vspace{0.3cm}

\noindent This gives the desired result for $x,y\leq \sqrt{t}$ (the exponential term is bounded for such $x$ and $y$). For $x,y\geq \sqrt{t}$, the upper bound is greater than one so the bound holds in this case as well. Let us now consider the case $x\leq \sqrt{t}$ and $y>\sqrt{t}$.

\vspace{0.3cm}

By Lemma 6.1, 

\begin{equation*}
\frac{\mathbb{P}_t^{x,y}(Y(s)>0 \ \mathrm{for} \ s\in[0,t])}{\mathbb{P}_t^{x,y}(Y(s)>0 \ \mathrm{for} \ s\in\lbrace 0,...,t\rbrace)}
\end{equation*}

\vspace{0.3cm}

\noindent is increasing in $x$ and $y$. Thus for $x\leq \sqrt{t}$ and $y>\sqrt{t}$, using \eqref{eq:exact} we see that

\begin{align*}
\mathbb{P}_t^{x,y}(B_0)&\leq \frac{\mathbb{P}_t^{x,y}(Y(s)>0 \ \mathrm{for} \ s\in[0,t]))}{\mathbb{P}_t^{x,\sqrt{t}}(Y(s)>0 \ \mathrm{for} \ s\in[0,t]))}\mathbb{P}_t^{x,\sqrt{t}}(B_0)\\
&\leq \frac{1-e^{-\frac{2xy}{t}}}{1-e^{-\frac{2x}{\sqrt{t}}}}C\frac{(1+x)(1+\sqrt{t})}{t}.
\end{align*}

\vspace{0.3cm}

\noindent One can then check that for $x\leq \sqrt{t}$ and $y>\sqrt{t}$,

\begin{equation*}
\frac{1-e^{-\frac{2xy}{t}}}{1-e^{-\frac{2x}{\sqrt{t}}}}\leq \frac{y}{\sqrt{t}}.
\end{equation*}

\vspace{0.3cm}

\noindent So we find that 

\begin{equation*}
\mathbb{P}_t^{x,y}(Y(s)>0 \ \mathrm{for} \ s\in\lbrace 0,...,t\rbrace)\leq C\frac{(1+x)(1+y)}{t}
\end{equation*}

\vspace{0.3cm}

\noindent for all $x,y\geq 0$. $_{_\square}$

\vspace{0.3cm}

We shall now briefly go over the final arguments in proving form of the lower order contributions to $m^\beta(t)$. We shall completely gloss over the technical details. The case of $\beta=\sqrt{2\log K}$ and $\beta>\sqrt{2\log K}$ need to be treated separately. We shall first consider $\beta>\sqrt{2\log K}$. 

\vspace{0.3cm}

\bf Lemma 6.3. \rm For $\beta>\sqrt{2\log K}$, there is a constant $C$ so that $m^\beta(t)\geq \sqrt{2\log K}t-\frac{3}{2\sqrt{2\log K}}\log t+C$.

\vspace{0.3cm}

\it Proof: \rm The first thing to note is that one can show that there is a constant $C'$ so that $m^\beta(t)\geq m^H(t)+C'$, where $m^H$ is the centering term in the $\beta=\infty$ case. Thus we only consider the $\beta=\infty$ case. We shall also write $m(t)=m^H(t)$ for this lemma. The bulk of the technical work in this lemma consists of showing that for any fixed $y_0$, $x\geq m(t)$, $y\geq y_0$ and large enough $r$ (which is considered fixed with respect to $t$), there is a constant $\tilde{C}$ (depending on $r$) so that

\begin{equation*}
\mathbb{E}_t^{x,y}\left(e^{\sum_{s=1}^t k_{t-s}(Y_s)}\right)\geq \tilde{C} K^t \mathbb{P}_t^{0,0}(Y(s)>0 \ \mathrm{for} \ s\in\lbrace r,...,t-r\rbrace).
\end{equation*}

\vspace{0.3cm}

To do this, one has to work a fair amount to identify the paths with significant weight in the Feynman-Kac formula and also show that the measure of this set of paths can be compared with $K^t \mathbb{P}_t^{0,0}(Y(s)>0 \ \mathrm{for} \ s\in\lbrace r,...,t-r\rbrace)$. 

\vspace{0.3cm}

Using \eqref{eq:ineq}, one can then argue that for some constant $\hat{C}$ (depending on $r$), 

\begin{equation*}
\mathbb{P}_t^{0,0}(Y(s)>0 \ \mathrm{for} \ s\in\lbrace r,...,t-r\rbrace)\geq \frac{\hat{C}}{t}.
\end{equation*}

\vspace{0.3cm}

These estimates then imply that for any $y_0$, $x\geq m(t)$, $U_t=1-G_t$, with $G_t$ given by the recursion relation with Heaviside initial data, one has

\begin{equation*}
U_t(x)\geq \int_{y_0}^\infty U_0(y)\frac{e^{-\frac{(x-y)^2}{2t}}}{\sqrt{2\pi t}}K^t \frac{C_1}{t}dy.
\end{equation*}

\vspace{0.3cm}

\noindent On the other hand, if one sets $x=\sqrt{2\log K}t-\frac{3}{3\sqrt{2\log K}}\log t+z_1$, where $|z_1|\leq C_2\sqrt{t}$ for some $C_2>0$, a quick calculation shows that for any fixed $y_0<0$

\begin{equation*}
\int_{y_0}^\infty U_0(y)\frac{e^{-\frac{(x-y)^2}{2t}}}{\sqrt{2\pi t}}dy\geq C_3 tK^{-t}e^{-\sqrt{2\log K}z_1}.
\end{equation*}

\vspace{0.3cm}

\noindent Let us now assume that for any fixed $z_1$, one can find a $t$ so that if we choose $x$ as above, $x>m(t)$. We note that this is equivalent to saying that for any constant $C$, one can find a $t$ so that $m(t)<\sqrt{2\log K}t-\frac{3}{2\sqrt{2\log K}}\log t+C$. So for such a $t$ it follows that $U_t(x)\geq C_4 e^{-\sqrt{2\log K}z_1}$. But with a suitable choice of $z_1$, this will imply that $U_t(x)>1$, which is impossible. So we conclude that for some constant $C$, 

\begin{equation*}
m(t)\geq \sqrt{2\log K}t-\frac{3}{2\sqrt{2\log K}}\log t+C
\end{equation*}

\vspace{0.3cm}

\noindent for all $t$. As we noted at the beginning, this implies that the same bound holds for $m^\beta(t)$. $_{_\square}$

\vspace{0.3cm}

\bf Lemma 6.4. \rm For $\beta>\sqrt{2\log K}$, there is a constant $C$ so that $m^\beta(t)\leq \sqrt{2\log K}t-\frac{3}{2\sqrt{2\log K}}\log t+C$.

\vspace{0.3cm}

\it Proof: \rm We consider again $U=1-G$ and define $U_0^*(x)=1$ for $x<0$ and $U_0^*(x)=U_0(x)$ for $x\geq 0$. We note that $U_0^*(x)\geq U_0(x)$ for all $x$ and $U_0^*(x)\geq U_0^H(x)$ where $U_0^H$ is the initial data in the Heaviside case, i.e., when $\beta=\infty$. If $U_t^*$ is then given by the recursion relation with initial data $U_0^*$, one can check that $U_t^*(x)\geq U_t(x)$ and $U_t^*(x)\geq U_t^H(x)$ for all $t$ and $x$. Moreover, $U_t^*$ is strictly decreasing so $m^*(t)=(U_t^*)^{-1}(\frac{1}{2})$ is well defined. Since $U_t^*$ is decreasing, $m^*(t)\geq m^\beta(t)$ so we only need to show the result for $m^*(t)$.  Let us write $m(t)$ for the centering term in the $\beta=\infty$ case and $k_s$ for the term in the exponential of the Feynman-Kac formula in the $\beta=\infty$ case. 

\vspace{0.3cm}

The majority of the technical work (which we shall skip) for this lemma goes into proving that

\begin{equation*}
\mathbb{E}_t^{x,y}\left(e^{\sum_{s=1}^t k_{t-s}(Y(s)}\right)\leq C'K^t\mathbb{P}_t^{\bar{z},\max(y,1)}(Y(s)>0 \ \mathrm{for} \ s\in\lbrace r,...,t-r\rbrace),
\end{equation*}

\vspace{0.3cm}

\noindent where $x\geq m(t)+1$, $y$ is arbitrary, $C'>0$ depending on $r$, $\bar{z}=x-\sqrt{2\log K}t+\frac{3}{2\sqrt{2\log K}}\log t+\tilde{C}$ for a suitable $\tilde{C}$ and $r$ is taken large enough (though fixed with respect to $t$).

\vspace{0.3cm}

Taking this result as a given, $U_t^*(x)\geq U_t^H(x)$ for all $x$ and $t$ then implies that $k^*_t(x)\leq k_t(x)$ for all $x$ and $t$. Thus we have the same upper bound for the expectation with $k^*$ instead of $k$. Plugging this into the Feynman-Kac representation for the recursion relation of $U_t^*$ and performing some estimation, one finds that there is a constant $\hat{C}$ so that for $x\geq m(t)+1$,

\begin{equation*}
U_t^*(x)\leq \hat{C} K^t \int_0^\infty U_0(y)\frac{1}{\sqrt{2\pi t}}e^{-\frac{(x-y)^2}{2t}}\mathbb{P}_t^{\bar{z},y}(Y(s)>0 \ \mathrm{for} \ s\in\lbrace r,...,t-r\rbrace)dy.
\end{equation*}

\vspace{0.3cm}

\noindent Writing $x=z_1+\sqrt{2\log K}t$ and using Lemma 6.2 one finds

\begin{align*}
U_t^*(x)&\leq \hat{C} e^{-\sqrt{2\log K}z_1} \int_0^\infty e^{\sqrt{2\log K}y}U_0(y)\frac{1}{\sqrt{2\pi t}}e^{-\frac{(z_1-y)^2}{2t}}\mathbb{P}_t^{\bar{z},y}(Y(s)>0 \ \mathrm{for} \ s\in\lbrace r,...,t-r\rbrace)dy\\
&\leq \hat{C}t^{-\frac{3}{2}}(1+\bar{z})e^{-\sqrt{2\log K}z_1}\int_0^\infty e^{\sqrt{2\log K}y}(1+y)U_0(y)dy. 
\end{align*}

\vspace{0.3cm}

Since $\beta>\sqrt{2\log K}$, the integral converges and we only care about the part depending on $t$, $z_1$ and $\bar{z}$. Noting that $\bar{z}=z_1-\frac{3}{2\sqrt{2\log K}}\log t+\tilde{C}$, we see that we can move the $t^{-\frac{3}{2}}$ into the exponential and get

\begin{equation*}
U_t^*(x)\leq \bar{C}(1+\bar{z})e^{-\sqrt{2\log K}\bar{z}}.
\end{equation*}

\vspace{0.3cm}

Using similar arguments to those in sections 4 and 5, one can show that $U_t(x+m^*(t))$ converges uniformly to $u=1-w$. We note that since $m^*(t)+D\geq m(t)+1$ for some constant $D$, we can set $x=m^*(t)+D$ in this estimate. Since $U_t^*(m^*(t)+D)$ converges to $u(D)>0$, the sequence is bounded from below by some positive number. The inequality above then implies that $\bar{z}=m^*(t)+D-\sqrt{2\log K}t+\frac{3}{2\sqrt{2\log K}}\log t+\tilde{C}$ must be bounded from above (otherwise the right side of the inequality could get arbitrarily close to zero). So we conclude that there is some constant $C$ so that 

\begin{equation*}
m^\beta(t)\leq m^*(t)\leq \sqrt{2\log K}t-\frac{3}{2\sqrt{2\log K}}\log t+C. _{_\square}
\end{equation*}

\vspace{0.3cm}

Combining these results shows that for $\beta>\sqrt{2\log K}$, 

\begin{equation*}
m^\beta(t)-\sqrt{2\log K}t+\frac{3}{2\sqrt{2\log K}}\log t
\end{equation*}

\vspace{0.3cm}

\noindent is bounded. We shall now proceed to the $\beta=\sqrt{2\log K}$ case. 

\vspace{0.3cm}

\bf Lemma 6.5. \rm Let $m:\Z_+\rightarrow \R$ satisfy $\frac{m(t)}{t}\rightarrow \sqrt{2\log K}$ as $t\rightarrow\infty$, $m(t)\geq \sqrt{2\log K}t-t^\delta$ and $m(t)-m(s)\geq \sqrt{2\log K}(t-s)-(t-s)^\delta-C_1$ for $s_0\leq s\leq t$ and $t$ chosen large enough. Moreover, assume that

\begin{equation*}
K^t\mathbf{E}^x(U_0(X(t)); X(s)>m(t-s)\ \mathrm{for} \ s\in\lbrace 0,...,t-s_0\rbrace)=D_t(z) v(z)
\end{equation*}

\vspace{0.3cm}

\noindent for some functions $v$ and $D_t$, where $z=x-m(t)$ and $D_t$ satisfies

\begin{equation*}
\liminf_{z\rightarrow\infty}\liminf_{t\rightarrow\infty} D_t(z)>0
\end{equation*}

\vspace{0.3cm}

\noindent and $s_0$ is such that for a fixed $t$, $m(s)$ is finite for $s\in\lbrace s_0,...,t\rbrace$. Then for $\beta=\sqrt{2\log K}$, $m(t)-m^\beta(t)$ is bounded for $t\geq s_0$. 

\vspace{0.3cm}

\it Proof: \rm  Consider $M_t=\max\lbrace |m(s)-m^\beta(s)|: s_0\leq s\leq t\rbrace$. The first thing to do in this lemma, that would require some work and whose proof we shall skip, is to show that under our assumptions concerning $m(t)$, 

\begin{equation*}
\frac{\mathbf{E}^x(U_0(X(t)); X(s)>m^\beta(t-s) \ \mathrm{for} \ s\in\lbrace 0,...,t-s_0\rbrace)}{\mathbf{E}^x(U_0(X(t)); X(s)>m(t-s) \ \mathrm{for} \ s\in\lbrace 0,...,t-s_0\rbrace)}
\end{equation*}

\vspace{0.3cm}

\noindent can be estimated from below in terms of a ratio of integrals concerning probabilities of the form $\mathbb{P}_s^{a,b}(Y(\tau)>0 \ \mathrm{for} \ \tau\in\lbrace 0,...,s\rbrace)$. Then using Lemma 6.2, one can show that for $c>0$ and some $N_c\geq 1$, 

\begin{equation*}
\frac{\mathbf{E}^x(U_0(X(t)); X(s)>m^\beta(t-s) \ \mathrm{for} \ s\in\lbrace 0,...,t-s_0\rbrace)}{\mathbf{E}^x(U_0(X(t)); X(s)>m(t-s) \ \mathrm{for} \ s\in\lbrace 0,...,t-s_0\rbrace)}\geq C\frac{y(t)}{1+y(t)+M_t}\frac{z'}{1+z'+M_t}e^{-cM_t}
\end{equation*}

\vspace{0.3cm}

\noindent for some $C>0$ if $x>m^\beta(t)+N_c$, where $y(t)$ is some sequence satisfying $y(t)\rightarrow \infty$ as $t\rightarrow\infty$ and $z'=x-m^\beta(t)-N_c$. We note that that since $U_0(x)>0$ for all $x$, then always

\begin{equation*}
K^t\mathbf{E}^x(U_0(X(t)); X(s)>m(t-s)\ \mathrm{for} \ s\in\lbrace 0,...,t-s_0\rbrace)>0
\end{equation*}

\vspace{0.3cm}

\noindent which implies that $v(z)>0$ for large enough $z$. Indeed taking $z$ large enough and then fixing it, our assumptions imply that for large enough $t$

\begin{equation*}
K^t\mathbf{E}^x(U_0(X(t));X(s)>m(t-s) \ \mathrm{for} \ s\in\lbrace 0,...,t-s_0\rbrace)\geq C(z),
\end{equation*}

\vspace{0.3cm}

\noindent where $C(z)>0$. Combining these estimates, we see that for large enough (fixed) $z>0$  and $t>0$

\begin{equation*}
K^t\mathbf{E}^x(U_0(X(t)); X(s)>m^\beta(t-s) \ \mathrm{for} \ s\in\lbrace 0,...,t-s_0\rbrace)\geq C(z)\frac{y(t)}{1+y(t)+M_t}\frac{z'}{1+z'+M_t}e^{-cM_t}.
\end{equation*}

\vspace{0.3cm}

Let us now show that $m(t)-m^\beta(t)$ is bounded. We begin by assuming that for some large fixed value of $t$, $m^\beta(t)\leq m(t)$ and set $z\geq N_c+1$. Then $z'\geq 1$ and the right side of the inequality above is at least

\begin{equation*}
C(z)\frac{1}{(2+M_t)^2}e^{-cM_t}\geq C(z)e^{-2cM_t}
\end{equation*}

\vspace{0.3cm}

\noindent for a large enough $M_t$ (if no such $t$ can be found, either $M_t$ is bounded or we can find one in the $m(t)\leq m^\beta(t)$ case we consider soon). So we conclude that under these assumptions 

\begin{equation*}
K^t\mathbf{E}^x(U_0(X(t)); X(s)>m^\beta(t-s) \ \mathrm{for} \ s\in\lbrace 0,...,t-s_0\rbrace)\geq C(z)e^{-2cM_t}.
\end{equation*}

\vspace{0.3cm}

\noindent The second thing that would require some work (which we shall not do) is to show that for $b<\sqrt{2\log K}$ and large enough (fixed) $t$ (for which $m^\beta(t)\leq m(t)$ is satisfied), one has

\begin{align*}
K^t\mathbf{E}^x&(U_0(X(t)); X(s)>m^\beta(t-s) \ \mathrm{for} \ s\in\lbrace 0,...,t-s_0\rbrace)\leq C_3e^{-bz'}
\end{align*}

\vspace{0.3cm}

\noindent for  $x\geq x_1=m^\beta(t)+N^b$ for large enough $N_b$. We recall that $x=m(t)+z=m^\beta(t)+N_c+z'$, so $z'=z+m(t)-m^\beta(t)-N_c$ and

\begin{align*}
C_3e^{-bz'}\leq C_4 e^{-bz}e^{-b(m(t)-m^\beta(t))}.
\end{align*}

\vspace{0.3cm}

\noindent Combining our estimates, we see that

\begin{equation*}
C(z)e^{-2cM_t}\leq C_4 e^{-bz}e^{-b(m(t)-m^\beta(t))}.
\end{equation*}

\vspace{0.3cm}

\noindent Now $z$ was large but fixed so if we set $c=\frac{1}{4}$ and $b=1$, then actually for such $t$, $m(t)-m^\beta(t)\leq \frac{1}{2}M_t+C_5$. One can than work a bit to show that if $m^\beta(t)\geq m(t)$, one can perform the same arguments with the roles of $m(t)$ and $m^\beta(t)$ switched.

\vspace{0.3cm}

Combining the estimates for $m(t)-m^\beta(t)$ and $m^\beta(t)-m(t)$, we see that$|m(t)-m^\beta(t)|\leq \frac{1}{2}M_t+C_6$. As $M_t$ is an increasing function of $t$, we see that for all $s\in\lbrace s_0,...,t\rbrace$, $|m(s)-m^\beta(s)|\leq \frac{1}{2}M_t+C_6$ which implies that $M_t\leq 2C_6$. Thus $M_t$ is bounded. $_{_\square}$

\vspace{0.3cm}

So all we have to do is to show that $m(t)=\sqrt{2\log K}t-\frac{1}{2\sqrt{2\log K}}\log t+\alpha_t$ satisfies the conditions of the previous lemma with some bounded $\alpha_t$. 

\vspace{0.3cm}

\bf Lemma 6.6. \rm For $m(t)=\sqrt{2\log K}t-\frac{3}{2\sqrt{2\log K}}\log t+b(t)$, with 

\begin{equation*}
b(t)=\frac{1}{\sqrt{2\log K}}\log\left(\int_0^\infty ye^{\sqrt{2\log K}y}U_0(y)e^{-\frac{y^2}{2t}}dy\right),
\end{equation*}

\vspace{0.3cm}

\noindent $m^\beta(t)-m(t)$ is bounded for $\beta=\sqrt{2\log K}$.

\vspace{0.3cm}

\it Proof: \rm We shall assume that when talking about $b$ and hence $m$, we are considering $t\geq 1$ so that everything is well defined. We begin by noting that $b(t)=\frac{\log t}{\sqrt{2\log K}}+\mathcal{O}(1)$. So indeed if we can show that $m(t)-m^\beta(t)$ is bounded we will have showed that $m^\beta(t)$ has the lower order behavior we claimed. The form of $b(t)$ also implies that $\frac{m(t)}{t}\rightarrow\sqrt{2\log K}$ as $t\rightarrow\infty$ and $m(t)\geq \sqrt{2\log K}t-t^\delta$ for large enough $t$. Also one can check that $m(t)-m(s)\geq \sqrt{2\log K}(t-s)-(t-s)^\delta-C_1$ for  $s\in\lbrace s_0,...,t\rbrace$ and some constants $C_1$ and $s_0$. So we only have to check the main condition of Lemma 6.5.

\vspace{0.3cm}

In this lemma, the majority of work (which we skip) goes into showing that for $x\geq m(t)$ and $z=x-m(t)$, one has

\begin{align*}
K^t\mathbf{E}^x&(U_0(X(t)); X(s)>m(t-s)\ \mathrm{for} \ s\in\lbrace 0,...,t-1\rbrace)&\\
&=D_t^1(z)K^t\int_{M_t}^\infty U_0(y)\frac{1}{\sqrt{2\pi t}}e^{-\frac{(x-y)^2}{2t}}\mathbb{P}_t^{z,y}\left(Y(s)>0 \  \mathrm{for} \ s\in\lbrace 0,...,t\rbrace\right)dy,
\end{align*}

\vspace{0.3cm}

\noindent where $D_t^1(z)\rightarrow 1$ if we first take $t\rightarrow\infty$ then $z\rightarrow\infty$. Also $M_t\rightarrow \infty$ although so slowly that it can be replaced by $-\infty$ without changing the asymptotic behavior of $D_t^1$. Due to the simple asymptotic form of our initial data and $m(t)$, one can show that the upper limit can be changed to $t^{\frac{1}{2}+\delta}$ for any $\delta>0$ without changing the asymptotic behavior of $D_t^1$. 

\vspace{0.3cm}

\noindent Now for any fixed $z\geq 1$ and $y\in[M_t,t^{\frac{1}{2}+\delta}]$, we can write

\begin{equation*}
\mathbb{P}_t^{z,y}(Y(s)>0 \ \mathrm{for} \ s\in\lbrace 0,...,t\rbrace)=F_t^1(z,y)\frac{zy}{t},
\end{equation*}

\vspace{0.3cm}

\noindent where by \eqref{eq:ineq}, $(F_t^1)_t$ is bounded from above and below by some positive constants. Also for the $z$, $y$ and $t$ we are interested in, we can write

\begin{align*}
e^{-\frac{(x-y)^2}{2t}}&=e^{\sqrt{2\log K}y-\frac{y^2}{2t}-t\log K}e^{-\frac{(z+m(t)-\sqrt{2\log K}t)^2}{2t}} e^{\frac{1}{t}(y-\sqrt{2\log K}t)(z+m(t)-\sqrt{2\log K}t)}\\
&=F_t^2(z,y) e^{\sqrt{2\log K}y-\frac{y^2}{2t}+t\log K-\sqrt{2\log K}(z+m(t))},
\end{align*}

\vspace{0.3cm}

\noindent where again $(F_t^2)_t$ is bounded from above and from below by some positive constants. Plugging these into our formula for the expectation, we find

\begin{align*}
K^t\mathbf{E}^x&(U_0(X(t)); X(s)>m(t-s)\ \mathrm{for} \ s\in\lbrace 0,...,t-1\rbrace)&\\
&=\tilde{D}_t(z)t^{-\frac{3}{2}}ze^{-\sqrt{2\log K}z}e^{-\sqrt{2\log K}(m(t)-\sqrt{2\log K}t)}\int_{M_t}^{t^{\frac{1}{2}+\delta}} U_0(y)ye^{\sqrt{2\log K}y}e^{-\frac{y^2}{2t}}dy,
\end{align*}

\vspace{0.3cm}

\noindent where $\tilde{D}_t$ satisfies $\liminf_{z\rightarrow\infty}\liminf_{t\rightarrow\infty}\tilde{D}_t(z)>0$. Using the definition of $m(t)$, a little more estimation then shows that actually

\begin{align*}
K^t\mathbf{E}^x&(U_0(X(t)); X(s)>m(t-s)\ \mathrm{for} \ s\in\lbrace 0,...,t-1\rbrace)&\\
&=D_t(z)t^{-\frac{3}{2}}ze^{-\sqrt{2\log K}z}e^{-\sqrt{2\log K}(m(t)-\sqrt{2\log K}t)}\int_{0}^{\infty} U_0(y)ye^{\sqrt{2\log K}y}e^{-\frac{y^2}{2t}}dy\\
&=D_t(z)ze^{-\sqrt{2\log K}z},
\end{align*}

\vspace{0.3cm}

\noindent where $\liminf_{z\rightarrow\infty}\liminf_{t\rightarrow\infty}D_t(z)>0$ and we can use the previous lemma to establish the desired result. $_{_\square}$

\section{Concluding remarks}

We have now concluded our main goal of showing that $G_t(x+m^\beta(t))$ converges to a traveling wave and we have seen the freezing transition at $\beta=\sqrt{2\log K}$. Moreover, we have seen that in the high temperature case, $m^\beta(t)$ is asymptotically linear in $t$ but at the critical point and at lower temperatures there are logarithmic corrections whose coefficients depend on whether or not we are precisely at the critical point. 

\vspace{0.3cm}

In this section, we shall discuss a bit further the relationship of these results to some of the problems mentioned in the introduction. First of all, let us see exactly the relationship between this problem and that of extreme value statistics (or the maximum of the branching random walk).

\vspace{0.3cm}

Consider the recursion relation with initial data $G_0$ and let $(X_k(t))_k$ denote the branching random walk. Consider now the functions

\begin{equation*}
F_t(x)=\mathbb{E}^x\left(\prod_{k=1}^{K^t} G_0(X_k(t))\right).
\end{equation*}

\vspace{0.3cm}

We can think of the branching random walk $(X_k(t+1))_k$ starting at $x$ as consisting of a jump from $x$ to $x+y$ and then $K$   branching random walks $(X_k^i(t))$ starting at $x+y$ which are all mutually independent, identically distributed and depend on $y$ only through the starting point. Independence and splitting the expectation into an expectation over $y$ and an expectation over the rest of the branching random walks gives

\vspace{0.3cm}

\begin{align*}
F_{t+1}(x)&=\int \frac{e^{-\frac{y^2}{2}}}{\sqrt{2\pi}}\prod_{i=1}^K \mathbb{E}^{x+y}\left(\prod_{k=1}^{K^t}U_0(X_k^i(t))\right)dy\\
&=\int \frac{e^{-\frac{y^2}{2}}}{\sqrt{2\pi}} F_t(x+y)^K dy.
\end{align*}

\vspace{0.3cm}

Since $F_0=G_0$, we see that $F_t=G_t$ for all $t$. Consider now the $\beta=\infty$ case. Then 

\begin{align*}
\prod_{k=1}^{K^t} G_0(X_k(t))&=\mathbf{1}\lbrace X_k(t)>0 \ \mathrm{for \ all} \ k\rbrace\\
&=\mathbf{1}\left\lbrace \min_k X_k(t)>0\right\rbrace.
\end{align*}

\vspace{0.3cm}

\noindent Since we are dealing with symmetric random variables, we see that this implies that

\begin{equation*}
G_t(x)=\mathbb{P}^0\left(\max_k X_k(t)\leq x\right).
\end{equation*}

\vspace{0.3cm}

So this is the explicit relationship between our problem in the $\beta=\infty$ case and the problem of extreme value statistics for certain logarithmically correlated random variables. As discussed in \cite{FB}, in case the $X_k$ were independent, the limiting distribution is well known and it is of the Gumbel form. In fact, the Gumbel distribution is the limit for a large class of dependent random variables as well. One specific feature of the Gumbel distribution is that at $\infty$ it behaves like $1-Ce^{-\beta x}$. We have seen that for $\beta\geq \sqrt{2\log K}$ we have logarithmically correlated random variables for which the limiting distribution of the maximum has a tail of the form $1-Cxe^{-\beta x}$ which means they are not in the Gumbel universality class. From the point of view of extreme value statistics, the interesting and probably difficult problem is to describe the universality class to which our random variables belong. Generalizing our recursion relation might be one way to get a hold of some other members in this class.

\vspace{0.3cm}

The other problem our recursion relation is related to was the problem of the multiplicative cascade measures on hypercubes. What we have now showed is that

\begin{equation*}
g_t(e^{-\beta x},\beta)=G_t^\beta(x+m^\beta(t))=\mathbb{E}(\exp(-e^{-\beta x}e^{-\beta m^\beta(t)}\mathcal{Z}(t)))
\end{equation*}

\vspace{0.3cm}

\noindent converges as $t\rightarrow \infty$. Here $g_t(u,\beta)$ is the Laplace transform of the random variable $M_t^\beta=e^{-\beta m^\beta(t)}\mathcal{Z}(t)$.  From our analysis, we see that $M_t^\beta$ is of the form $A_t t^\alpha e^{-\beta c(\beta)t}\mathcal{Z}(t)$, where $\alpha\in\lbrace 0,\frac{1}{2},\frac{3}{2}\rbrace$ and $A_t$ is bounded. Since the Laplace transforms converge to some function which is continuous at $0$ ($g_t(u,\beta)\rightarrow w_\beta(-\frac{1}{\beta}\log u)$), we see that $M_t^\beta$ converge in distribution to some random variable, say $M^\beta$. So we have showed that after a deterministic normalization of the multiplicativa cascade measures, their total masses converge.

\vspace{0.3cm}

From Lemma 5.5, $m^\beta(t+1)-m^\beta(t)\rightarrow c(\beta)$ so this convergence in distribution and decomposing $\mathcal{Z}(t+1)=e^{-\beta V}\sum_{k=1}^K \mathcal{Z}^{k}(t)$ imply that $M^\beta$ satisfies the following equation in distribution

\begin{equation*}
M^\beta=e^{-\beta (V+c(\beta))}\sum_{k=1}^K M^{\beta,k},
\end{equation*}

\vspace{0.3cm}

\noindent where $V$ is a standard Gaussian and $M^{\beta,k}$ are independent copies of $M^\beta$ and also independent of $V$. This equation is of the form that is studied in \cite{DL}. While their main results are about existence of solutions to such equations, which is not of direct interest to us since we have a solution, they do use a construction that is of interest in our case. Consider the $\beta\geq \sqrt{2\log K}$ case and let $g(u,\beta)=\lim_t g_t(u,\beta)$. Since for $\beta \geq \sqrt{2\log K}$, $w_\beta=w_{\sqrt{2\log K}}$, we have for $\beta\geq \sqrt{2\log K}$

\begin{equation*}
g(u,\beta)=g\left(u^{\frac{\sqrt{2 \log K}}{\beta}},\sqrt{2\log K}\right).
\end{equation*}

\vspace{0.3cm}

So we see that the Laplace transforms of the limits have a rather simple $\beta$ dependence. A similar phenomenon occurs in \cite{DL}. Applying their argument in our case implies that $M^\beta$ is given by composing a certain stable process with the random variable $M^{\sqrt{2\log K}}$. 

\vspace{0.3cm}

The point of these remarks is to possibly give some tools or approaches to questions such as almost sure convergence of $M^{\beta}_t$ or even the existence and structure of the limiting measure. 

\vspace{0.3cm}

\subsection*{Acknowledgments:}

I wish to thank Antti Kupiainen for introducing me to this subject and constantly finding time for useful discussions and advice. I also wish to thank Greg Lawler for suggesting the approach to Lemma 6.2 and the referee for helpful comments and suggestions for references. Finally I wish to thank the Academy of Finland for financial support.

\bibliography{final}
\bibliographystyle{plain}

\end{document}